\documentclass[aps,prd,amsmath,%
               superscriptaddress,onecolumn,12pt,tightenlines]{revtex4}

\usepackage{graphicx}

\newcommand{\ie}{\textit{i.e.}}
\newcommand{\eg}{\textit{e.g.}}

\newcommand{\gae}{%
  \ensuremath{\lower 2pt \hbox{%
    $\, \buildrel {\scriptstyle >}\over {\scriptstyle \sim}\,$}%
    }%
  }
\newcommand{\lae}{%
  \ensuremath{\lower 2pt \hbox{%
    $\, \buildrel {\scriptstyle <}\over {\scriptstyle \sim}\,$}%
    }%
  }

\newcommand{\refeqn}[2][eqn:]{Eqn.~(\ref{#1#2})}

\newcommand{\reffig}[2][fig:]{Figure~\ref{#1#2}}
\newcommand{\refFig}[2][fig:]{Figure~\ref{#1#2}}
\newcommand{\refsec}[2][sec:]{Section~\ref{#1#2}} 

\newcommand{\erf}{\mathop{\textrm{erf}}}
\newcommand{\rmd}{\ensuremath{\mathrm{d}}}  

\newcommand{\dRdE}{\ensuremath{\mathcal{R}}}
\newcommand{\vmin}{\ensuremath{v_\textrm{min}}}
\newcommand{\vmp}{\ensuremath{\overline{v}_0}}
\newcommand{\vobs}{\ensuremath{v_\textrm{obs}}}
\newcommand{\bvobs}{\ensuremath{\mathbf{v}_\textrm{obs}}}
\newcommand{\vesc}{\ensuremath{v_\textrm{esc}}}
\newcommand{\Nesc}{\ensuremath{N_\textrm{esc}}}
\newcommand{\bu}{\ensuremath{\mathbf{u}}}  
\newcommand{\bv}{\ensuremath{\mathbf{v}}}  
\newcommand{\bV}{\ensuremath{\mathbf{V}}}  
\newcommand{\eone}{\ensuremath{\hat{\boldsymbol{\varepsilon}}_1}}  
\newcommand{\etwo}{\ensuremath{\hat{\boldsymbol{\varepsilon}}_2}}  
\newcommand{\egen}{\ensuremath{\hat{\boldsymbol{\varepsilon}}_i}}  
\newcommand{\peone}{\ensuremath{\hat{\mathbf{e}}_1}}  
\newcommand{\petwo}{\ensuremath{\hat{\mathbf{e}}_2}}  
\newcommand{\Eco}{\ensuremath{E_\textrm{co}}}         
\newcommand{\Ecomin}{\ensuremath{E_\textrm{co,min}}}  
\newcommand{\Ecomax}{\ensuremath{E_\textrm{co,max}}}  
\newcommand{\SHM}{\ensuremath{\textrm{SHM}}}   
\newcommand{\Sgr}{\ensuremath{\textrm{Sgr}}}   
\newcommand{\Str}{\ensuremath{\textrm{str}}}   
\newcommand{\orderof}[1]{\ensuremath{\mathcal{O}(#1)}}

\newcommand{\insertfig}[1]{%
    \includegraphics[keepaspectratio,width=1.00\columnwidth,
                     height=0.35\textheight]{#1}
}

\begin{document}


\preprint{MCTP-06-16}


\title{Annual Modulation of Dark Matter in the Presence of Streams}

\author{Chris Savage}
\email[]{cmsavage@umich.edu}
\affiliation{
 Michigan Center for Theoretical Physics,
 Department of Physics,
 University of Michigan,
 Ann Arbor, MI 48109}

\author{Katherine Freese}
\email[]{ktfreese@umich.edu}
\affiliation{
 Michigan Center for Theoretical Physics,
 Department of Physics,
 University of Michigan,
 Ann Arbor, MI 48109}

\author{Paolo Gondolo}
\email[]{paulo@physics.utah.edu}
\affiliation{
 Physics Department,
 University of Utah,
 Salt Lake City, UT 84112}

\date{\today}


\begin{abstract}
  
  In addition to a smooth component of WIMP dark matter in galaxies,
  there may be streams of material; the effects of WIMP streams on
  direct detection experiments is examined in this paper.  The
  contribution to the count rate due to the stream cuts off at some
  characteristic energy. Near this cutoff energy, the stream
  contribution to the annual modulation of recoils in the detector is
  comparable to that of the thermalized halo, even if the stream
  represents only a small portion ($\sim$5\% or less) of the local halo
  density.  Consequently the total
  modulation may be quite different than would be expected for the
  standard halo model alone: it may not be cosine-like and can peak at a
  different date than expected.  The effects of speed, direction,
  density, and velocity dispersion of a stream on the modulation are
  examined.  We describe how the observation of a modulation can be used
  to determine these stream parameters.  Alternatively, the presence of
  a dropoff in the recoil spectrum can be used to determine the WIMP
  mass if the stream speed is known.  The annual modulation of the
  cutoff energy together with the annual modulation of the overall
  signal provide a ``smoking gun'' for WIMP detection.

\end{abstract}

\maketitle


\section{\label{sec:Intro} Introduction}

The Milky Way, along with other galaxies, is well known to be
encompassed in a massive dark matter halo of unknown composition.  
Leading candidates for this dark matter are Weakly
Interacting Massive Particles (WIMPs), a generic class of particles that
includes the lightest supersymmetric particle.
Numerous collaborations worldwide have been
searching for these particles. Direct detection experiments attempt to
observe the nuclear recoil caused by these dark matter particles
interacting with nuclei in the detectors.  These experiments include
DAMA/NaI \cite{Bernabei:2003za},
DAMA/LIBRA \cite{Bernabei:2005ki},
NAIAD \cite{Alner:2005kt},
CDMS \cite{Akerib:2005kh},
EDELWEISS \cite{Sanglard:2005we},
ZEPLIN \cite{Alner:2005pa},
XENON \cite{Aprile:2004ey},
DRIFT \cite{Alner:2004cw,Alner:2005},
CRESST \cite{Angloher:2002in,Angloher:2004tr},
SIMPLE \cite{Girard:2005pt},
PICASSO \cite{Barnabe-Heider:2003cq},
COUPP \cite{Bolte:2005fm},
and many others.
It is well known that the count rate in WIMP direct detection
experiments will experience an annual modulation
\cite{Drukier:1986tm, Freese:1987wu} as a result of the motion of the
Earth around the Sun: the relative velocity of the detector with respect
to the WIMPs depends on the time of year.  

The actual signals in a dark matter detector, including the modulation,
depend on the distribution of WIMPs in the galaxy.  It is commonly
believed that a majority of the WIMPs have become thermalized into a
smooth halo distribution.  In the simplest model, the halo is a
spherically symmetric, non-rotating isothermal sphere of WIMPs
\cite{Freese:1987wu}.  The annual modulation of non-standard halos, such
as anisotropic models, is discussed in Refs.~\cite{Green:2000jg,
Green:2003yh,Ullio:2000bf,Fornengo:2003fm,Copi:2002hm}.

However, galaxy formation is a continual process, with new material
still being accreted through, \eg, absorption of dwarf galaxies such as
the Sagittarius dwarf galaxy \cite{Yanny:2003zu,Newberg:2003cu,
Majewski:2003ux}.  The result is that the halo can contain a non-trivial
amount of substructure such as clumps and tidal streams of material.
The presence of such substructure is supported by $N$-body simulations
of galaxy formation (see, \eg, Refs.~\cite{Klypin:1999uc,Moore:1999nt}).
Unlike the virialized component of the halo, a clump or stream of
material would result in a ``cold'' flow of WIMPs through a detector:
the velocity dispersion is small relative to the typical speed with
respect to the Earth, so that the WIMPs are
incident from nearly the same direction and with nearly the same speed.
Alternative models of halo formation, such as the late-infall model
\cite{Gunn:1972sv,Fillmore:1984wk,Bertschinger:1900nj} (more recently
examined by Sikivie and others \cite{Sikivie:1992bk,Sikivie:1996nn,
Sikivie:1997ng,Sikivie:1999jv,Tremaine:1998nk,Natarajan:2005fh}),
also predict cold flows of
dark matter.  Any such streaming of WIMPs (we will henceforth use
``stream'' to imply any cold flow) will yield a significantly different
modulation effect than that due to a smooth halo.
A stream actually provides two types of modulation: a modulation in the
overall signal and a modulation in some cutoff energy above which counts
due to the stream are not observed.  Together, these two types of
modulation can yield a ``smoking gun'' for WIMPs.
The modulation in the presence of one or more streams is discussed in
Refs.~\cite{Gelmini:2000dm,Stiff:2001dq,Ling:2004aj,Freese:2003na,
Freese:2003tt}.

Here, we examine how various parameters describing a stream affect the
modulation signal; the parameters examined are the stream direction,
speed, density, and dispersion.  We expect the stream density (and hence
contribution to the recoil rate) to be small, $\orderof{\textrm{few\%}}$
that of the isothermal Halo, and yet find that the stream can have
significant effects on the annual modulation.  We note the following
interesting modifications to the overall annual modulation when we add
the effects of the stream into the isothermal halo:
\begin{itemize}
  \item The combined modulation generally does not have a cosinuoidal
        variation with time, even if the modulation of each individual
        component does;
  \item The combined modulation need not be time-symmetric, even if each
        individual components is; and
  \item The minimum and maximum recoil rates need not occur 0.5 years
        apart.
  \item Near the cutoff energy of the stream (above which it no longer
        contributes), the stream contribution to the annual modulation 
        is comparable to that of the halo, even for a 2\% stream
        density (relative to the background halo). The drastic effects
        of the stream near the cutoff energy can be seen in
        \reffig{Components} (for the modulation shown at 39 keV).
  \item The annual modulation of the cutoff energy together with the
        annual modulation of the overall signal provide a ``smoking
        gun'' for WIMP detection.
\end{itemize}
As a consequence of the second-to-last point, it is likely that
the existence of a stream will be identified near its cutoff energy.
Since a stream's effects are mild except near $E_c$ (and essentially
non-existent well above $E_c$), the presence of a stream should not
interfere with using the modulation to describe the background
distribution, SHM or otherwise.  

In \refsec{DMDetection}, we will begin by reviewing direct detection 
techniques for WIMPs and the time-dependent WIMP recoil rate in an
experimental detector.  We will describe the Standard Halo Model (SHM)
and examine the behavior of the modulation signals in this model in
\refsec{SHM}.  We will then examine how a dark matter stream affects
this signal, first with the Sagittarius (Sgr) stream
(\refsec{SgrStream}), then generalizing to other streams
(\refsec{Streams}).  In \refsec{Streams}, we also briefly outline how
various stream parameters can be extracted from a modulation signal.
Our results are summarized in \refsec{Summary}.

\section{\label{sec:DMDetection} Dark Matter Detection}

WIMP direct detection experiments seek to measure the energy deposited
when a WIMP interacts with a nucleus in the detector \cite{Goodman:1984dc}.
If a WIMP of mass $m$ scatters elastically from a nucleus of mass $M$,
it will deposit a recoil energy $E = (\mu^2v^2/M)(1-\cos\theta)$,
where $\mu \equiv m M/ (m + M)$ is the reduced mass, $v$ is the speed
of the WIMP relative to the nucleus, and $\theta$ is the scattering
angle in the center of mass frame.  The differential recoil rate per
unit detector mass for a WIMP mass $m$, typically given in units of
counts/kg/day/keV, can be written as:
\begin{equation} \label{eqn:dRdE}
  \dRdE \equiv
  \frac{\rmd R}{\rmd E} = \frac{\sigma(q)}{2 m \mu^2}\, \rho\, \eta(E,t)
\end{equation}
where $q = \sqrt{2 M E}$ is the nucleus recoil momentum, $\sigma(q)$ is
the WIMP-nucleus cross-section, $\rho$ is the local WIMP density,
and information about the WIMP velocity distribution is encoded into the
mean inverse speed $\eta(E,t)$,
\begin{equation} \label{eqn:eta}  
  \eta(E,t) = \int_{u>\vmin} \frac{f(\bu,t)}{u} \, \rmd^3u \, .
\end{equation}
Here 
\begin{equation} \label{eqn:vmin}
  \vmin = \sqrt{\frac{M E}{2\mu^2}}
\end{equation}
represents the minimum WIMP velocity that can result in a recoil energy
$E$ and $f({\bf u},t)$ is the (time-dependent) distribution of WIMP
velocities ${\bf u}$ relative to the detector.  For distinguishable WIMP
populations (such as multiple streams), $\rho \eta \to
\sum_i \rho_i \eta_i$ in \refeqn{dRdE}, where $\rho_i$ and $\eta_i$ are
the local density and mean inverse speed, respectively, of some WIMP
population indexed by $i$.

In this paper, the use of the term ``rate'' will refer to the
differential rate $\dRdE$.  Dates will be given as fractions of a
calendar year (\ie\ from January 1).  For example, June 1 will be given
as $t = 0.415$ years.  For illustrative purposes, we will take
$\sigma_p = 10^{-42}$ cm$^2$ (defined in the following section) and use
germanium 
as our detector element.  In addition, as an example 
we take the WIMP mass to be $m = 60$ GeV and will use a local density of
$\rho = 0.3$ GeV/cm$^3$ for the smooth WIMP halo (streams will be in
addition to the smooth halo, so the total local density can exceed this
value).

\subsection{\label{sec:CS} Cross-Section}

The cross-section for WIMP interactions is given by
\begin{equation} \label{eqn:CS}
  \sigma(q) = \sigma_0 \, F^2(q) \, ,
\end{equation}
where $\sigma_0$ is the zero-momentum WIMP-nuclear cross-section and
$F(q)$ is the nuclear form factor, normalized to $F(0) = 1$; a
description of these form factors may be found in
Refs.~\cite{SmithLewin,Jungman:1995df}.  We will assume a purely scalar
interaction, although the true interaction is likely to have both scalar
and spin-dependent components.  The results of this paper are not
qualitatively affected by the choice of interaction: the primary
difference would be a change in the recoil rates by an overall factor;
the shape of the modulation would not change.

For purely scalar interactions,
\begin{equation} \label{eqn:scalar}
  \sigma_0 = \frac{4 \mu^2}{\pi} [Zf_p + (A-Z)f_n]^2 \, .
\end{equation}
Here $Z$ is the number of protons, $A-Z$ is the number of neutrons,
and $f_p$ and $f_n$ are the WIMP couplings to the proton and nucleon,
respectively.  In most instances, $f_n \sim f_p$; the WIMP-nucleus
cross-section can then be given in terms of the WIMP-proton
cross-section as a result of \refeqn{scalar}:
\begin{equation} \label{eqn:SICSproton}
  \sigma_0 = \sigma_p \left( \frac{\mu}{\mu_p} \right)^2 A^2
\end{equation}
where the $\mu_p$ is the WIMP-proton reduced mass, and $A$ is
the atomic mass of the target nucleus.
Again, for illustrative purposes, we have chosen $\sigma_p = 10^{-42}$
cm$^2$, a germanium ($Z = 32$, $A \approx 73$) detector, and a WIMP mass
of $m = 60$ GeV.

\subsection{\label{sec:VelocityDist} Velocity Distribution}

We will be examining two sets of WIMP populations: (1) a smooth
galactic halo component, see \refsec{SHM}, and (2) streams.  We study
the Sagittarius stream for illustration in \refsec{SgrStream} and
generalize to all streams in \refsec{Streams}.  For both halo and
stream components, we will use a Maxwellian distribution, characterized
by a velocity dispersion $\sigma_v$, to describe the WIMP speeds, and
we will allow for the distribution to be truncated at some escape
velocity $\vesc$,
\begin{equation} \label{eqn:Maxwellian}
  \widetilde{f}(\bv) =
    \begin{cases}
      \frac{1}{\Nesc} \left( \frac{3}{2 \pi \sigma_v^2} \right)^{3/2}
        \, e^{-3\bv^2\!/2\sigma_v^2} , 
        & \textrm{for} \,\, |\bv| < \vesc  \\
      0 , & \textrm{otherwise}.
    \end{cases}
\end{equation}
Here
\begin{equation} \label{eqn:Nesc}
  \Nesc = \erf(z) - 2 z \exp(-z^2) / \pi^{1/2} ,   
\end{equation}
with $z \equiv \vesc/\vmp$, is a normalization factor.  The most
probable speed,
\begin{equation} \label{eqn:vmp}
  \vmp = \sqrt{2/3} \, \sigma_v ,
\end{equation}
is used to generate unitless parameters such as $z$.
For distributions without an escape velocity ($\vesc \to \infty$),
$\Nesc = 1$.

The WIMP component (halo or stream) often exhibits a bulk motion
relative to us, so that
\begin{equation} \label{eqn:vdist}
  f(\bu) = \widetilde{f}(\bvobs + \bu) \, ,
\end{equation}
where $\bvobs$ is the motion of the observer relative to the rest frame
of the WIMP component described by \refeqn{Maxwellian}; this motion will
be discussed in the following section.  For such a velocity
distribution, the mean inverse speed, \refeqn{eta}, becomes
\begin{equation} \label{eqn:eta2}
  \eta(E,t) =
    \begin{cases}
      \frac{1}{\vmp y} \, ,
        & \textrm{for} \,\, z<y, \, x<|y\!-\!z| \\
      \frac{1}{2 \Nesc \vmp y}
        \left[
          \erf(x\!+\!y) - \erf(x\!-\!y) - \frac{4}{\sqrt{\pi}} y e^{-z^2}
        \right] \, ,
        & \textrm{for} \,\, z>y, \, x<|y\!-\!z| \\
      \frac{1}{2 \Nesc \vmp y}
        \left[
          \erf(z) - \erf(x\!-\!y) - \frac{2}{\sqrt{\pi}} (y\!+\!z\!-\!x) e^{-z^2}
        \right] \, ,
        & \textrm{for} \,\, |y\!-\!z|<x<y\!+\!z \\
      0 \, ,
        & \textrm{for} \,\, y\!+\!z<x
    \end{cases}
\end{equation}
where 
\begin{equation} \label{eqn:x}
  x \equiv \vmin/\vmp \, ,
\end{equation}
$y \equiv \vobs/\vmp$, and
$z \equiv \vesc/\vmp$.  Here, we use the common notational convention of
representing 3-vectors in bold and the magnitude of a vector in the
non-bold equivalent, \eg\ $\vobs \equiv |\bvobs|$.


\textit{Isothermal (Standard) Halo Model.}
For WIMPs in the Milky Way halo, the most frequently employed WIMP
velocity distribution is that of a simple non-rotating isothermal sphere
\cite{Freese:1987wu}, also referred to as the Standard Halo Model (SHM).
Typical parameters of the Maxwellian distribution for our location in
the Milky Way are $\sigma_\SHM$ = 270 km/s and $\vesc$ = 650 km/s, the
latter being the speed necessary to escape the Milky Way (WIMPs with
speeds in excess of this would have escaped the galaxy, hence the
truncation of the distribution in \refeqn{Maxwellian}).
More complex models, allowing for \eg\ anisotropy and triaxiality, may
better match the actual dark matter halo; see Ref.~\cite{Green:2002ht}
and references therein.  Such models do not qualitatively effect the
results of this paper, as any smooth halo will give the same general
behavior as exhibited by the SHM (non-smooth components, \eg\ clumps,
will result in streaming WIMPs, the focus of this paper).  Then, for
simplicity, we will take the SHM to describe the halo.
The SHM will be discussed in \refsec{SHM}.


\textit{Sagittarius stream.}
The Sagittarius (Sgr) dwarf galaxy is being absorbed by the Milky Way
and has a tidally stripped tail passing through the disk very near to us
\cite{Yanny:2003zu,Newberg:2003cu,Majewski:2003ux}.  This tail might
provide a stream of WIMPs observable in dark matter detectors.  While
we are interested in streams in general, we will use this Sgr stream to
illustrate how various stream properties affect the the signal in dark
matter detection.  In most cases, we will assume a dispersion of
$\sigma_\Sgr$ = 25 km/s and an untruncated Maxwellian
($\vesc \to \infty$) for this stream.  The Sgr stream will be examined
in \refsec{SgrStream}.


\textit{General streams.}
In addition to the Sgr stream, there may be tidal streams from other
dwarf galaxies being absorbed by the Milky Way or streams arising from
late infall of dark matter.  The dispersion $\sigma_\Str$ can vary, but
will generally be much smaller than the observer's velocity $\vobs$ in
\refeqn{vdist}.  The Maxwellian for these streams will also be
untruncated ($\vesc \to \infty$).  General streams will be explored in
\refsec{Streams}.

\subsection{\label{sec:EarthMotion} Motion of the Earth}

Due to the motion of the Earth around the Sun, $\bvobs$ is time
dependent: $\bvobs = \bvobs(t)$.  We write this in terms of the Earth's
velocity $\bV_\oplus$ relative to the Sun as
\begin{equation} \label{eqn:vobs}
  \bvobs(t) = \bv_\odot
              + V_\oplus \left[
                  \eone \cos{\omega(t-t_1)} + \etwo \sin{\omega(t-t_1)}
                \right] \, ,
\end{equation}
where $\omega = 2\pi$/year, $\bv_\odot$ is the Sun's motion relative to
the WIMP component's rest frame, $V_\oplus = 29.8$ km/s is the Earth's
orbital speed, and $\eone$ and $\etwo$ are the directions of the Earth's
velocity at times $t_1$ and $t_1+0.25$ years, respectively.  Henceforth,
as mentioned previously, all times will be given as fractions of a
calendar year (\ie\ from January 1).
With this form, we have neglected the ellipticity of the Earth's orbit,
although the ellipticity is small and, if accounted for, would give only
negligible changes in the results of this paper (see
Refs.~\cite{Green:2003yh,SmithLewin} for more detailed expressions).
For clarity, we have used explicit velocity vectors rather than the
position vectors $\peone$ and $\petwo$ used in
Refs.~\cite{Gelmini:2000dm,Freese:2003tt} and elsewhere (the position
vectors are more easily generalized to an elliptical orbit); the two
bases are related by $\eone = -\petwo$ and $\etwo = \peone$.

In Galactic coordinates, where $\hat{\mathbf{x}}$ is the direction to
the Galactic Center, $\hat{\mathbf{y}}$ the direction of disk rotation,
and $\hat{\mathbf{z}}$ the North Galactic Pole,
\begin{eqnarray}
  \label{eqn:eone}
    \eone &=& (0.9931, 0.1170, -0.01032) \, , \\
  \label{eqn:etwo}
    \etwo &=& (-0.0670, 0.4927, -0.8676) \, ,
\end{eqnarray}
where we have taken $\eone$ and $\etwo$ to be the direction of the
Earth's motion at the Spring equinox and Summer solstice, respectively,
with $t_1 = 0.218$ the fraction of the year before the Spring equinox
(March 21).

The time dependence of $\vobs$ simplifies to the form
\begin{equation} \label{eqn:vobst}
  \vobs(t) = \sqrt{v_\odot^2 + V_\oplus^2
                   + 2\,b\,v_\odot V_\oplus \cos{\omega(t-t_c)}} \, ,
\end{equation}
where $b = \sqrt{b_1^2 + b_2^2}$ for $b_i \equiv \egen \cdot
\hat{\bv}_\odot$, and $t_c$ is the solution of
\begin{equation} \label{eqn:tc}
  \cos{\omega (t_c - t_1)} = \frac{b_1}{\sqrt{b_1^2 + b_2^2}} \, , \quad
  \sin{\omega (t_c - t_1)} = \frac{b_2}{\sqrt{b_1^2 + b_2^2}} \, .
\end{equation}
The parameter $b$ is a geometrical factor relating to the orthogonality
of $\bv_\odot$ with the Earth's orbital plane: $b = \sin\lambda_\odot$,
where $\lambda_\odot$ is the angle between $\bv_\odot$ and the normal to
the orbital plane, with a maximum value of 1 when $\bv_\odot$ is in the
orbital plane and a minimum value of 0 when $\bv_\odot$ is completely
orthogonal to the plane.  We define the \textbf{characteristic time} as
\begin{description}
  \setlength{\itemindent}{\listparindent}  
  \item[$t_c$ =] the time of year at which $\vobs$
        is maximized.
\end{description}
In the typical case in which the Earth's orbital speed is significantly
smaller than the net motion of the WIMP population (or
$V_\oplus \ll v_\odot$) so that relative changes in $\vobs(t)$ are
small, as with the SHM and (observable) streams,
\begin{equation} \label{eqn:vobstapprox}
  \vobs(t) \approx v_\odot \left[
                     1 + b \frac{V_\oplus}{v_\odot} \cos{\omega(t-t_c)}
                   \right] \, .
\end{equation}


\textit{Standard Halo Model.}
Unlike the Galactic disk (along with the Sun), the halo has essentially
no rotation; the motion of the Sun relative to this stationary halo is
\begin{equation} \label{eqn:vsunSHM}
  \bv_{\odot,\SHM} = \bv_{\textrm{LSR}} + \bv_{\odot,\textrm{pec}}
    \, ,
\end{equation}
where $\bv_{\textrm{LSR}} = (0,220,0)$ km/s is the motion of the
Local Standard of Rest and $\bv_{\odot,\textrm{pec}} = (10,13,7)$ km/s
is the Sun's peculiar velocity.  The characteristic time is
$t_{c,\SHM}$ = 0.415 (June 1) and the geometrical parameter $b$ has the
value of 0.49.  The SHM will be discussed in \refsec{SHM}.


\textit{Sagittarius stream.}
The Sgr stream is moving at approximately $300 \pm 90$ km/s relative to
the galactic rest frame with a 3-velocity in this frame of
\begin{equation} \label{eqn:vSgr}
  \bv_\Sgr = 300 \, \textrm{km/s}\, \times \, (0,0.233,-0.970) \, .
\end{equation}
The derivation of this velocity, described in
Refs.~\cite{Majewski:2003ux,Freese:2003tt}, allows for significant
uncertainties; however, as we will primarily use this stream simply as
an example of streams in general, we will ignore these uncertainties
and use only the stated values.  The velocity of the stream relative to
the Sun is then
\begin{equation} \label{eqn:vsunSgr}
  \bv_{\odot,\Sgr} = \bv_{\textrm{LSR}} + \bv_{\odot,\textrm{pec}}
                     - \bv_{\Sgr} \, ,
\end{equation}
with $\bv_{\textrm{LSR}}$ and $\bv_{\odot,\textrm{pec}}$ as given above,
yielding $v_{\odot,\Sgr}$ = 340 km/s.  The characteristic time for this
WIMP population is then $t_{c,\Sgr}$ = 0.991 (Dec 28); $b$ in
\refeqn{vobst} is equal to 0.53.  The Sgr stream will be examined in
\refsec{SgrStream}.


\textit{General streams.}
For a general stream, we allow both the direction and magnitude of the
velocity $\bv_{\odot,\Str}$ to vary.  The characteristic time
$t_{c,\Str}$ can take any date and $b$ can take any value from 0 to 1;
both parameters are completely specified by the stream direction
$\hat{\bv}_{\odot,\Str}$.  General streams will be explored in
\refsec{Streams}.

\subsection{\label{sec:Modulation} Annual Modulation}

It is well known that the count rate in WIMP detectors will experience
an annual modulation as a result of the motion of the Earth around the
Sun \cite{Drukier:1986tm,Freese:1987wu}.  In some cases, but not all,
the count rate (\refeqn{dRdE}) has an approximate time dependence
\begin{equation} \label{eqn:dRdEapprox}
  \dRdE(t)
  \approx \dRdE(t_c + 0.25)
          \left[
            1 + \orderof{1} \, b \frac{V_\oplus}{v_\odot} \cos{\omega(t-t_c)}
          \right] \, ,
\end{equation}
where the $\orderof{1}$ factor may be positive or negative.  For a
cosine-like modulation, $t_c + 0.25$ years corresponds to the time of
year at which the rate is approximately average; again, times are given
as fractions of a year.  Since we shall show that, in many cases, the
modulation is decidedly not cosine-like, we divide the recoil rate
$\dRdE(t)$ more generally into a time-averaged component $\dRdE_0$ and
time-residual (modulation) component $\dRdE_m(t)$:
\begin{equation} \label{eqn:modulation}
  \dRdE(t) = \dRdE_0 + \dRdE_m(t) = \dRdE_0 [1 + \chi_m(t)] \, ,
\end{equation}
where $\chi_m(t) \equiv \frac{\dRdE_m(t)}{\dRdE_0}$ is the fractional
(relative) modulation amplitude; by definition, $\dRdE_m(t)$ averages to
zero over one year.

Typical WIMP recoil energies (on the order of 10's of keV) are
comparable to that due to various background sources; experimental
detectors thus have the difficult problem of needing to distinguish
between background events and actual WIMP recoils, where the number
of background events is much larger than the expected number of WIMP
events.  Presumably, however, such background does not exhibit the
same time dependent behavior that is expected of WIMPs.  For large
detectors that can obtain high statistics, the WIMP modulation may
be detectable even without precisely knowing the backgrounds present
in the data; in this case, $\dRdE_m$ is therefore determined without
knowing $\dRdE_0$.  DAMA used a large NaI detector to search for such
a signal and found an annual modulation \cite{Bernabei:2003za}.
This signal is incompatible with the null results of other experiments,
such as CDMS, under the conventional assumptions of elastic,
spin-independent interactions and an isothermal halo
\cite{Abrams:2002nb} (with some exceptions \cite{Gondolo:2005hh}).
These results, however, may be compatible for spin-dependent
\cite{Ullio:2000bv,Savage:2004fn} or inelastic interactions
\cite{Smith:2001hy,Bernabei:2002qr,Tucker-Smith:2004jv}.
Interpretation of the DAMA signal for non-standard halos is discussed in
Refs.~\cite{Belli:1999nz,Belli:2002yt}; DAMA also examined the
possibility of streams in Ref.~\cite{Bernabei:2006ya}.  The presence of
an energy cutoff due to a stream in this modulation signal would provide
a ``smoking gun'' for WIMPs.

The primary purpose of this paper is to examine the annual modulation
due to WIMP streams and constrast this modulation with that due to the
smooth galactic component.  We will begin by examining the smooth
galactic halo component in \refsec{SHM}.  We will then compare how a
stream modulation differs from the galactic component and study how
various stream parameters affect the modulation, first using the
Sagittarius stream for illustrative purposes in \refsec{SgrStream} and
then with more general streams in \refsec{Streams}.

\section{\label{sec:SHM} Isothermal (Standard) Halo Model}

As noted in the previous section, the most frequently employed WIMP
velocity distribution for the Milky Way halo is that of a simple
isothermal sphere \cite{Freese:1987wu}, also referred to as the Standard
Halo Model (SHM), given by \refeqn{Maxwellian} with $\sigma_\SHM$ = 270
km/s and $\vesc$ = 650 km/s.  For these parameters, previously discussed
in \refsec{VelocityDist}, and the parameters given in
\refsec{EarthMotion} (we will also take the local WIMP density for this
smooth halo component alone to be $\rho_\SHM$ = 0.3 GeV/cm$^3$ in this
and all following sections), the modulation of the rate is well
approximated by a cosine over a large range of energies, with a relative
residual rate (see
\refeqn{modulation})
\begin{equation} \label{eqn:chiSHM}
  \chi_m(x,t) \approx \cos{\omega(t-t_c)} \times
    \begin{cases}
      -0.034 \, (1 - \frac{x^2}{x_p^2}) \, ,
        & \textrm{for} \,\, x<x_p, \\ 
      0.014 \, (\frac{x}{x_p} - 1) (\frac{x}{x_p} + 3.7) \, ,
        & \textrm{for} \,\, x_p<x\lae z, 
    \end{cases}
\end{equation}
where $t_c =0.415$ (June 1), $z = 2.95$, and $x_p = 0.89$ is the value
of $x$ at which the phase of the modulation reverses (recall
$x \equiv \vmin/\vmp$ with $\vmin \propto \sqrt{E}$).  For $x \gae z$,
the relative size of the modulation grows and the shape becomes
non- cosine-like; however, the overall rate becomes highly suppressed.
From \refeqn{chiSHM}, we can see two features:
\begin{enumerate}
  \item The modulation is only a few percent of the average count rate.
        Thus, a large number of events are required to observe a
        modulation of the rate in a detector.
  \item For $x<x_p$ (low recoil energies), the cosine is multiplied by
        a negative factor: the rate is \textit{minimized} at a time
        $t_c$.  For $x>x_p$ (high recoil energies), the reverse is
        true: the rate is \textit{maximized} at time $t_c$.
        See Ref.~\cite{Lewis:2003bv} for a discussion of this phase
        reversal.
\end{enumerate}

\begin{figure}
  \insertfig{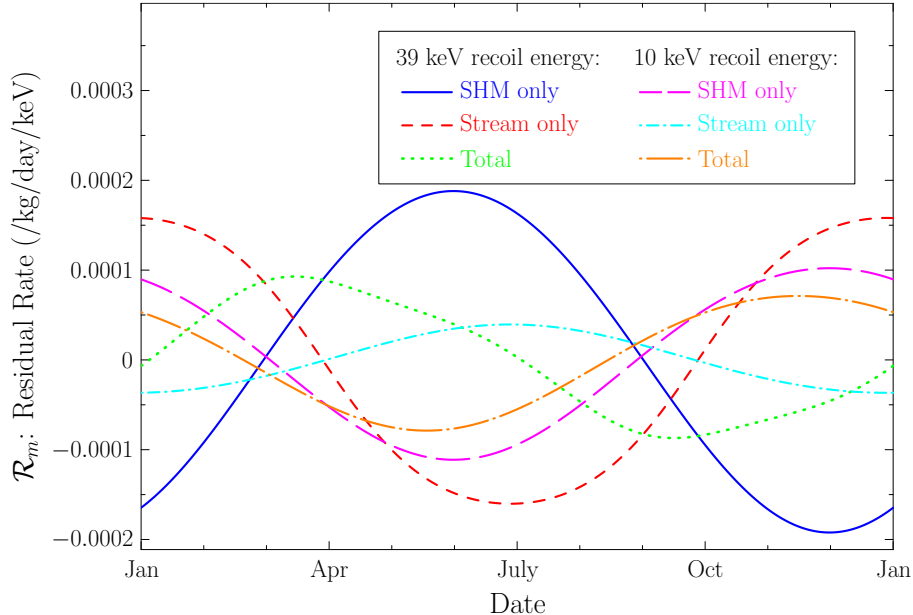}
  \caption[Components]{
    The modulation due to the Standard Halo Model (SHM) and Sagittarius
    (Sgr) stream separately, as well as the combined modulation, for a
    Germanium detector.  Each component is shown at a recoil energies of
    10 keV and 39 keV.  The WIMP mass is assumed to be 60 GeV and the
    Sgr density is taken to be 5\% of the SHM density.
    }
  \label{fig:Components}
\end{figure}

The detector type (composed of an element with mass $M$) and WIMP mass
$m$ arise in \refeqn{chiSHM} only through a constant scaling of the
parameter $x = \frac{\vmin}{\vmp}$ with the square root of the recoil
energy $E$ via \refeqn{vmin}.  Thus, the modulation behavior of
\refeqn{chiSHM} and the two features listed above are expected to be
observed in any detector, for any WIMP mass, but at different energy
scales.  For the sake of illustration, we will assume a WIMP mass $m=60$
GeV and a Germanium ($A \approx 73$) detector.  In this case $x=0.25
\sqrt{E/1\textrm{keV}}$; the parameter value of $x_p$ = 0.89 at which
the phase of the modulation reverses in \refeqn{chiSHM} corresponds to
an energy $E_p = 13$ keV.  Then for $E<13$ keV, there will be a
cosine-like modulation in the recoil rate at the few percent level,
minimized around June 1; see \reffig{Components} for an example.  For
$13$ keV $<E<110$ keV, a cosine-like modulation will occur, but will be
maximized around June 1; again see \reffig{Components}.  For
$E \gae 110$ keV, the overall rate becomes insignificantly small
(compared to lower energies).  Experiments such as
CDMS \cite{Akerib:2005kh} and EDELWEISS \cite{Sanglard:2005we},
which use Germanium detectors, have thresholds low enough to potentially
see this phase reversal, but are not expected to observe enough events
to discern the small (few percent) modulation effect.
Experiments using other elements in their detectors, such as
ZEPLIN \cite{Alner:2005pa} and XENON \cite{Aprile:2004ey} (using Xenon),
likewise could see such a reversal if the exposure is sufficient to
detect the modulation effect.

\begin{figure}
  \insertfig{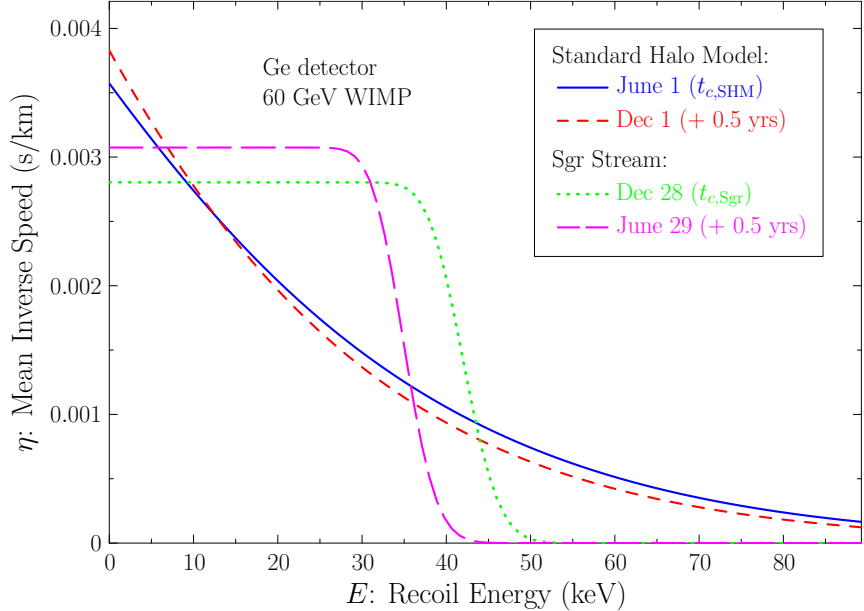}
  \caption[$\eta$]{
    The mean inverse speed $\eta$, given by \refeqn{eta}, for the SHM
    and Sgr stream at $t_c$ and $t_c$ + 0.5 years, where $t_c$ is
    different for the two WIMP populations.
    The characteristic time $t_c$ is the time of year at which the
    Earth is moving fastest relative to the given WIMP population
    (see \refsec{EarthMotion}).
    For the SHM, $\eta$ at $t_{c,\SHM}$ (June 1) is slightly larger at
    higher recoil energies and smaller at lower recoil energies than at
    $t_{c,\SHM}$ + 0.5 years (Dec 1); the change in the phase occurs at
    an energy $E_{p,\SHM}$ = 13 keV.  In the SHM,
    the fractional change in $\eta$ is
    of order $\frac{V_\oplus}{v_{\odot,\SHM}}$ (only a few percent)
    for all energies.  For the Sgr stream at $t_{c,\Sgr}$ (Dec 28),
    $\eta$ is flat up to a cutoff near the characteristic energy
    $E_{c,\Sgr}$ = 39 keV (see \refeqn{EcSgr}); the cutoff is softened
    by the velocity dispersion of WIMPs in the stream.  At $t_{c,\Sgr}$
    + 0.5 years (June 29), the cutoff energy has decreased to around 35
    keV due to the lower maximum WIMP velocities, but the plateau height
    at energies below this has increased relative to $t_{c,\Sgr}$.  For
    $\eta$ at energies below the cutoff, the relative change over the
    year is also of order $\frac{V_\oplus}{v_{\odot,\SHM}}$.
    However, due to the shift in the cutoff, the change in $\eta$ is
    \textit{not} suppressed by this velocity factor over a small range
    of recoil energies near 39 keV; the change here is \orderof{1}
    relative to the average at that energy.
    }
  \label{fig:Eta}
\end{figure}

Both features, the small modulation amplitude and phase change, arise
predominantly from the mean inverse speed factor $\eta$ in
\refeqn{dRdE}; $\eta$ is shown in \reffig{Eta} at both $t_c$ and
$t_c$ + 0.5 years.  At $t_c$, the Earth is moving the fastest relative
to the SHM, shifting the WIMP velocity distribution, \refeqn{vdist}, to
higher velocities compared to $t_c$ + 0.5 years (when the Earth is
moving slowest).  The total recoil rate (over all energies) is highest
at $t_c$, but the shift to higher velocities depletes the number of low
velocity WIMPs and low energy recoils.  However, the small change in our
velocity throughout the year relative to our velocity through the
SHM, along with the large velocity dispersion of WIMPs (comparable to
our motion through the halo), yield only small changes in $\eta$;
the fractional change is of order $\frac{V_\oplus}{v_{\odot,\SHM}}$ for
all energies.  The large velocity dispersion reduces the modulation
effect because, even in our frame, there are significant numbers of
WIMPs incident on Earth from all directions (but somewhat more likely
from the ``forward'' direction): in some directions, the WIMP velocity
is actually reduced at $t_c$ and increased at $t_c$ + 0.5 years.
The geometrical factor $b$ in \refeqn{vobst} (relating to the angle of
the Sun's motion through the galaxy with the Earth's orbital plane),
here having the value 0.49, further suppresses the modulation amplitude.

\begin{figure}
  \insertfig{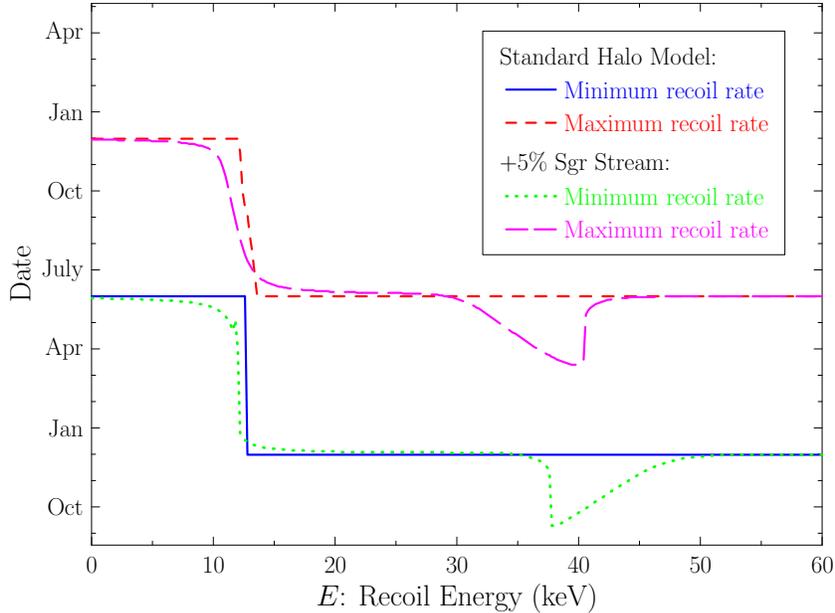}
  \caption[Dates of minimum/maximum recoil rate]{
    The dates at which the minimum and maximum recoil rates occur for
    various recoil energies.
    For the SHM alone, as demonstrated in \reffig{Components}, the
    modulation is symmetric about $t_{c,\SHM}$ (June 1), with a minimum
    at this time for energies below $E_{p,\SHM}$ = 13 keV and a maximum
    here for energies above this; the other extremum is always at
    $t_{c,\SHM}$ + 0.5 years (Dec 1).
    The Sgr stream component alone is symmetric about $t_{c,\Sgr}$
    (Dec 28); but when the SHM and Sgr stream are combined (shown here
    with a 5\% stream density relative to the SHM density), the total
    modulation becomes asymmetric.
    The asymmetry is apparent at 30-50 keV, where the maxima and minima
    are not 0.5 years apart.
    }
  \label{fig:Dates}
\end{figure}

As is apparent from \refeqn{chiSHM}, the modulation is symmetric in time
about $t_{c}$ (June 1) for the SHM.  The rate is always at an extremum
on this date, with the other extremum 0.5 years later (Dec 1), as seen
in \reffig{Dates}.  We shall see in the next section that there is no
symmetry when additional WIMP populations, such as streams, are present
in the halo: the date of the rate extrema changes with energy and the
minimum and maximum rate need not occur 0.5 years apart.

\section{\label{sec:SgrStream} Sagittarius Stream}

Beyond the smooth SHM dark matter distribution, the galaxy contains
additional structure, such as streams arising from late infall of dark
matter or tidal disruption of dwarf galaxies being absorbed by the Milky
Way.  While we are interested in examining streams in general, in this
section we will use the Sagittarius (Sgr) stream \cite{Yanny:2003zu,
Newberg:2003cu,Majewski:2003ux} as an example to illustrate how various
stream parameters affect the modulation signal in a detector.
As discussed previously, the Sgr stream is a stream of WIMPs
associated with a tidally stripped tail of the Sgr dwarf galaxy that is
currently being absorbed by our galaxy.  This tidal tail passes through
the Milky Way's disk very near to us, so the WIMPs in the Sgr stream are
potentially observable in direct detection experiments.
Detection of WIMPs from the Sgr stream has been discussed in
Refs.~\cite{Freese:2003na,Freese:2003tt}; here, we expand upon those
discussions.  We will discuss the dependence of the annual modulation in
the count rate on the recoil energy, the binning of the recoil energy,
the stream density, and the stream dispersion.  In \refsec{Streams}, we
will examine how the results illustrated with the Sgr stream can change
with more general streams.  From the parameters discussed in
\refsec{VelocityDist} and \refsec{EarthMotion}, the characteristic time
for this WIMP population, \ie\ the date at which we are moving fastest
relative to the stream, is $t_c$ = 0.991 (Dec 28).  The geometric
parameter $b$ in \refeqn{vobst}, associated with the angle of the stream
relative to the Earth's orbital plane, is equal to 0.53.

We first examine the basic behavior of a stream by neglecting the
velocity dispersion ($\sigma_v=0$). In this case, the mean inverse speed
$\eta$, \refeqn{eta}, has the constant value $\frac{1}{\vobs(t)}$
up to a \textit{cutoff energy} $\Eco(t)$,
\begin{equation} \label{eqn:etaSgr}
  \eta(E,t) = \frac{1}{\vobs(t)} \theta\left[ \Eco(t) - E \right] \, ,
\end{equation}
where $\theta$ is the Heaviside step function.  The cutoff energy
corresponds to $\vmin = \vobs(t)$ (see \refeqn{vmin}), so
\begin{equation} \label{eqn:EcoSgr}
  \Eco(t) 
          = E_c \left[ 1 + A_c \cos{\omega(t-t_c)} \right] \, ,
\end{equation}
where
\begin{equation} \label{eqn:AcSgr}
  A_c \equiv \frac{2 \, b \, v_{\odot,\Sgr} V_\oplus}
                  {v_{\odot,\Sgr}^2 + V_\oplus^2} .
\end{equation}
We define a \textbf{characteristic energy}
\begin{equation} \label{eqn:EcSgr}
  E_c \equiv \frac{2 \mu^2}{M} \langle \left[ \vobs(t) \right]^2 \rangle
      = \frac{2 \mu^2}{M} \left( v_{\odot,\Sgr}^2 + V_\oplus^2 \right)
        \, .
\end{equation}
The relevant velocities have been defined in \refsec{EarthMotion} (see,
\eg, Eqns.~(\ref{eqn:vSgr}) and~(\ref{eqn:vsunSgr})).
The characteristic energy $E_c$ is also the average cutoff energy for
the case of zero velocity dispersion discussed here, but the definition
in \refeqn{EcSgr} may still be used when a velocity dispersion is
included ($\sigma_v \neq 0$; see below), in which case there is no hard
cutoff energy at $\Eco(t)$.
While this definition can also be used to define an $E_c$ for any WIMP
population, $E_c$ is not a useful quantity for a WIMP population with
a large velocity dispersion ($\sigma_v \gae v_\odot$), such as the SHM,
as there is no associated rapid drop off in the count rates near that
energy such as with a stream (which has $\sigma_v \ll v_\odot$).  Hence,
we will only use the characteristic energy of streams in our discussion.
We wish to emphasize the importance of the step function in
\refeqn{etaSgr}: the presence of the step and the fact that its
position varies with time have key consequences that will be seen in
the discussion that follows.

As with the SHM modulation discussed in the previous section, the
modulation from the stream (arising from \refeqn{etaSgr}) occurs for
any detector type, composed of an element with mass $M$, and any WIMP
mass $m$.  However, the characteristic energy $E_c$ given by
\refeqn{EcSgr} and, thus, the location of the step as given by
\refeqn{EcoSgr} depend explicitly on $M$ and $m$ (recall $\mu \equiv
M m / (M+m)$ is the reduced mass).  The choice of $M$ and $m$ do not
change the qualitative behavior of the modulation, only the energy
scales at which various effects occur.  For illustrative purposes, we
take a $60$ GeV WIMP and a Germanium detector, for which $E_c$ = 39 keV
for the Sgr stream; our results will apply to other detectors and WIMP
masses, just at different energies.

For the dispersionless case,
\begin{equation} \label{eqn:chiSgr}
  \chi_m(E,t) \approx -0.046 \, \cos{\omega(t-t_c)} \, ,
                      \quad\quad \textrm{for} \,\, E < \Ecomin
\end{equation}
and there are no recoils for $E > \Ecomax$, where $\Ecomin$ and
$\Ecomax$ are the minimum and maximum cutoff energies, occurring at
$t_c$ + 0.5 years and $t_c$, respectively.  Below $\Ecomin$, the
relative variation in the recoil rate is of order 
$\frac{V_\oplus}{v_{\odot,\Sgr}}$ and the modulation of the rate is
cosine-like with a minimum at $t_c$; this behavior is similar to that
seen in the SHM for energies below the phase reversal energy $E_p$.
For $E > \Eco(t)$, the WIMPs in the stream are not moving sufficiently
fast to produce a recoil of energy $E$.  However, for
$\Ecomin < E < \Ecomax$, there are times during the year at which
$E > \Eco(t)$ ($\dRdE=0$) and there are times at which $E < \Eco(t)$
($\dRdE>0$).  In this case, the size of the modulation is $\orderof{1}$
relative to the average recoil rate due to the step in \refeqn{etaSgr},
much larger than the $\orderof{\frac{V_\oplus}{v_{\odot,\Sgr}}}$ effect
arising from the time dependence in $\frac{1}{\vobs(t)}$ that is
apparent in \refeqn{vobstapprox}; the modulation in this energy range
can be quite large and non- cosine-like.

The behavior of \refeqn{etaSgr} approximately holds for non-zero
velocity dispersion $\sigma_v$ as well, provided the dispersion is not
significantly larger than the changes in $\vobs(t)$ (\ie\ $\sigma_v \lae
b V_\oplus$), although the cutoff softens for non-zero $\sigma_v$.
For illustrative purposes, we take the velocity dispersion of
\refeqn{Maxwellian} to be $\sigma_v$ = 25 km/s for the Sgr stream
(corresponding to $\vmp$ = 20 km/s from \refeqn{vmp}),
although we will examine variations in this
parameter in \refsec{Dispersion}.  For this case, $\eta$ is shown in
\reffig{Eta} at both $t_c$ (Dec 28) and $t_c$ + 0.5 years (June 29).
In the figure, the softened step function is apparent, with a cutoff
around $E_c$ = 39 keV.  The height of the step and the cutoff energy
vary with time.  At $t_c$, the Earth is moving the fastest relative to
the stream.  This leads to a larger range of recoil energies as opposed
to $t_c$ + 0.5 years, when the Earth is moving the slowest, so the step
occurs at a larger energy at $t_c$ ($\sim$43 keV) than half a year later
($\sim$35 keV).  The higher WIMP velocities, by spreading the recoils
over a larger energy range, lead to less recoils at any given energy and
a lower step height.  The relative variation in the step height is
of order $b\,\frac{V_\oplus}{v_{\odot,\Sgr}}$ ($\sim$5\%), similar to
the relative variation in $\eta$ for the SHM (also a few percent).
However, the variation around the step is unlike any seen in the SHM:
$\eta$ goes from nearly zero at $t_c$ + 0.5 years to essentially the
full step height at $t_c$.  Just as with the dispersionless case
previously discussed, the dominant contribution to the modulation
in this energy range is the shift on and off the step, \textit{not} the
variation in the step height itself.  Thus, the relative variation in
the recoil rate (recall $\dRdE(t) \propto \eta(t)$) is not suppressed by 
the above velocity factor and $\chi_m \sim \orderof{1}$.  This is a
large amplification in the signal and could potentially yield an
observable signal for the Sgr stream at a comparable level to the SHM,
even if the Sgr stream is significantly less dense.

To examine the recoil rate for the Sgr stream and compare to the SHM,
we must assume a local density $\rho_\Sgr$ for the stream.  For
illustrative purposes, we will assume a stream density 5\% that of the
SHM ($\rho_\Sgr = 0.05 \rho_\SHM$).  We note this is optimistic and that
the local Sgr density is likely at most a few percent
\cite{Freese:2003tt}.  However, 5\% is not unreasonable for other
structure in the dark matter distribution, such as clumps
\cite{Stiff:2001dq}.  The various effects we examine still mainly apply
for lower densities, just to smaller degrees.  We will, however, examine
variations in the density in \refsec{Density}.

We will take $\rho_\Sgr = 0.05 \rho_\SHM$ and $\sigma_v = $ 25 km/s
to be our fiducial values in
Sections~\ref{sec:Components}-\ref{sec:RecoilEnergyRange}
and then we will vary them in Sections~\ref{sec:Density}
\&~\ref{sec:Dispersion}, respectively.  
We will examine the dependence of the modulation in the
count rate as a function of the recoil energy, the binning of the
recoil energy, the stream density, and the stream dispersion.
We wish to reiterate the following definitions, which play an essential
role in the remaining discussions:
\begin{description}
  \setlength{\itemindent}{\listparindent}  
  \item[the characteristic time $t_c$]is the time of year at which
        Earth is moving fastest relative to some WIMP population,
        formally defined by \refeqn{tc}, and
  \item[the characteristic energy $E_c$] is the approximate energy at
        which a rapid drop in the recoil rate occurs for streams,
        formally defined by \refeqn{EcSgr}.
\end{description}

We also reiterate that, for illustrative purposes, we take a $60$ GeV
WIMP and a Germanium detector, for which {\boldmath$E_c$}
\textbf{= 39 keV} for the Sgr stream; our results will apply to other
detectors and WIMP masses, just at different energies.

\subsection{\label{sec:Components} Contrasting and Summing the Halo and
            Stream Modulations}

Here we will compare the modulation due to a stream, using
the Sgr stream as an example, with the modulation due to the SHM as
discussed in \refsec{SHM} and we will examine the total modulation
(sum of the SHM and stream components) that a detector will observe.
Figures~\ref{fig:Components}-\ref{fig:Dates} demonstrate out results.

For a 5\% Sgr stream with $\sigma_v$ = 25 km/s, the residual rate
(modulation) is shown in \reffig{Components} at recoil energies of
10 keV and 39 keV.  At 10 keV, the modulation is essentially a cosine
minimized on December 28 ($t_{c,\Sgr}$) and maximized on June 29.
This behavior is similar to the low energy behavior of the SHM, although
the SHM has a different $t_c$ (June 1) and, hence, does not peak at the
same time.  At 39 keV, near the characteristic energy $E_{c,\Sgr}$ for
the Sgr stream, we see the modulation becomes extremely large for the
stream, with an amplitude nearly four times larger than at 10 keV, even
though the nuclear form factor in \refeqn{CS} generally causes the
recoil rate to \textit{decrease} at higher energies.  This amplification
in the amplitude is due to the $\eta$ step edge crossing discussed
previously.
The phase at 39 keV is reversed from that at 10 keV: the maximum now
occurs on December 28 ($t_{c,\Sgr}$) rather than on June 29.  However,
unlike the SHM, the stream modulation is not cosine-like at all
energies.  While this particular instance may look somewhat cosine-like,
most energies in the cutoff range around $E_c$ are decidedly not
cosine-like; a smaller $\sigma_v$ would also make the modulation at
$E_{c,\Sgr}$ less cosinusoidal (we shall see this in
\refsec{Dispersion}).

In general, even if the WIMP population components (\eg\ SHM and Sgr
stream) individually produce cosinusoidal modulations, the total
modulation seen in detectors is not cosinusoidal.  At 10 keV, the total
modulation shown in \reffig{Components} does look somewhat cosinusoidal;
this, however, is a coincidence due to the nearly six month difference
between the $t_c$'s of the two components.
At $E_{c,\Sgr}$ = 39 keV, the total modulation is clearly not
time-symmetric.

While the SHM and Sgr stream always have their maxima or minima at their
respective $t_c$'s for any energy (but possibly changing between the
maximum and minimum at some specific energy), the combined modulation
has extrema occurring on dates that vary with energy.  In
\reffig{Dates}, the SHM can be seen to always have the extrema on June 1
($t_{c,\SHM}$) and December 1, but with the extrema reversing at an
energy $E_{p,\SHM}$ = 13 keV.  With a 5\% stream in addition to the SHM,
the combined modulation is mostly the same as for the SHM alone except
near $E_{p,\SHM}$ and $E_{c,\Sgr}$.  At low energies ($E \lae 8$ keV),
the SHM has the most significant contribution to the modulation and the
extrema occur on nearly the same dates as the SHM alone.  At
intermediate energies (15 keV $\lae E \lae$ 30 keV), the SHM is again
the most significant component and the extrema of the total modulation
are near that of the SHM's.  Above 45 keV (above the maximum cutoff
energy $\Ecomax$ for the Sgr stream), there is essentially no
contribution from the stream, so the extrema match the SHM.  In the
regions near $E_{p,\SHM}$ and $E_{c,Sgr}$, however, the dates at which
the maximum and minimum count rates occur are different for the combined
SHM \& Sgr stream modulation than they would be for the SHM alone.
The SHM modulation amplitude varies smoothly
across zero at $E_{p,\SHM}$ and is small at nearby energies, as can be
seen in \reffig{Eta}.  The Sgr stream modulation amplitude does not vary
significantly at these energies, so the stream contributes significantly
to the modulation and leads to a smooth variation in the extrema dates.
Around and between $\Ecomin$ and $\Ecomax$ for the stream (33-45 keV),
the total modulation fluctuates wildly due to the cutoff in $\eta$.  The
modulation here is noticeably asymmetric in \reffig{Dates}, as the
extrema are not 0.5 years apart.

As noted above, several properties can emerge from multiple components
that are not present in the individual components:
\begin{itemize}
  \item The combined modulation is generally non-cosinusoidal, even if
        the modulation of each individual component is;
  \item The combined modulation need not be time-symmetric, even if each
        individual components is; and
  \item The minimum and maximum recoil rates need not occur 0.5 years
        apart.
\end{itemize}
These three effects, if observed, are potentially evidence for some
structure (streams) in the local WIMP population.  These effects will be
apparent in the following sections, as we examine the behavior of the
modulation for different recoil energies, recoil energy binning, stream
density, and stream dispersion.  For these sections, we will be
examining only the combined modulation of the SHM and Sgr stream (not
individually).

\subsection{\label{sec:RecoilEnergy} Recoil Energy}

\begin{figure}
  \insertfig{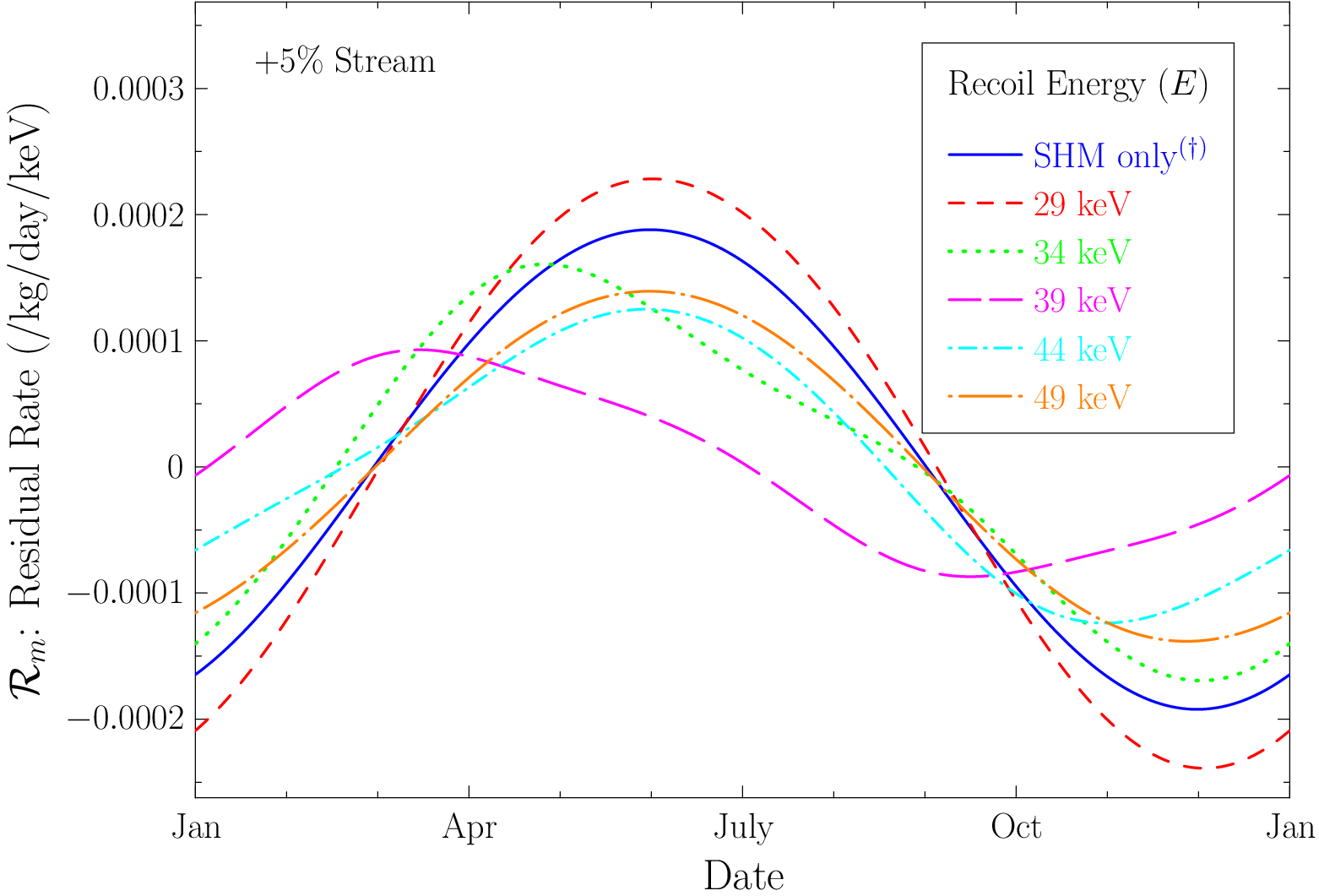}
  \caption[Recoil Energy]{
    \textit{Dependence on recoil energy.}
    The solid (blue) line shows the modulation due to only the SHM
    at a recoil energy of 39 keV $(\dag)$; the remaining lines include
    a 5\% Sgr stream in addition to the SHM at various recoil energies.
    The shape of the modulation is dependent on the recoil energy and
    clearly exhibits non- cosine-like behavior over a range of energies.
    The contribution from the stream falls off at energies above 39 keV,
    the characteristic energy $E_{c,\Sgr}$ for the Sgr stream (see
    \reffig{Eta}).
    }
  \label{fig:RecoilEnergy}
\end{figure}

Here we discuss the dependence of the total modulation on the recoil
energy.  As seen in \reffig{RecoilEnergy}, the shape of the modulation
is highly dependent on this energy.  In this figure, we have chosen to
show the total modulation at representative energies below, at, and
above 39 keV, the characteristic energy of the Sgr stream for 60 GeV
WIMPs in a Germanium detector.  For comparison, the cosine-like
contribution from the SHM alone is shown at 39 keV (solid/blue line).

Well below the characteristic energy of the stream, both the SHM and
stream modulations are cosine-like, although peaking at different times.
In \reffig{RecoilEnergy}, we see that the combination is approximately
cosine-like at 29 keV, a consequence of the two separate modulations
being nearly in phase with each other (both the SHM and Sgr stream peak
in the summer at this energy).  For the SHM (and other WIMP
populations), the cosine-like modulation comes from the expansion of the
SHM's speed relative to the Earth $\vobs(t)$ in the powers of
$\frac{V_\oplus}{v_\odot} \ll 1$ as given in \refeqn{vobstapprox}.  This
equation has corrections of $\orderof{\frac{V_\oplus^2}{v_\odot^2}}$.
Although not noticeable in the figure, the combined SHM + Sgr stream
modulation at 29 keV differs from a true cosine by a larger amount than
these $\orderof{\frac{V_\oplus^2}{v_\odot^2}}$ corrections.  A stream
with a different $t_c$ would yield a modulation that does not appear as
cosine-like.

At 39 keV ($E_{c,Sgr}$), the $\orderof{1}$ relative modulation of the
stream leads to a contribution to the modulation by the stream
comparable in magnitude to the contribution from the SHM, shifting the
peak date by nearly three months and greatly decreasing the modulation
amplitude relative to the SHM dominated behavior at 29 keV or SHM
component of the modulation alone at 39 keV.  This is a highly
significant result: even though the stream is much less dense than the
SHM, its modulation is just as large as that of the SHM here.  The
stream's effect on the total modulation is likely to be seen here well
before other energies.  The asymmetry in the modulation is apparent at
this energy.  The contribution to the total modulation from the stream
is also significant at energies near to $E_{c,\Sgr}$, as can be seen at
34 keV and 44 keV (the latter being more apparent near the minima in the
figure).

At 49 keV, above $\Ecomax$ for the stream, there is essentially no
contribution from the stream and the total modulation assumes the
cosine-like form of the SHM.  The amplitude is reduced from the case at
29 keV, which had contributions from the stream, and is also slightly
smaller than the SHM component alone at 39 keV, due to the form factor
in \refeqn{CS} that lowers the count rate at higher energies.

A stream can have a significant effect on the modulation near the
characteristic energy $E_c$ even if the stream is far less dense than
the smooth background halo distribution (here, the SHM).  The stream
can significantly change the peak date and the shape of the modulation
(becoming quite non-cosinusoidal) near this energy.  Away from this
energy, however, a low density stream has only a mild impact, if any, on
the modulation.  The question of how well experiments are able to
distinguish the modulation behavior near $E_c$, due to the limited
energy resolution in a detector, is addressed in the following section.

\subsection{\label{sec:RecoilEnergyRange} Recoil Energy Range (Binning)}

\begin{figure}
  \insertfig{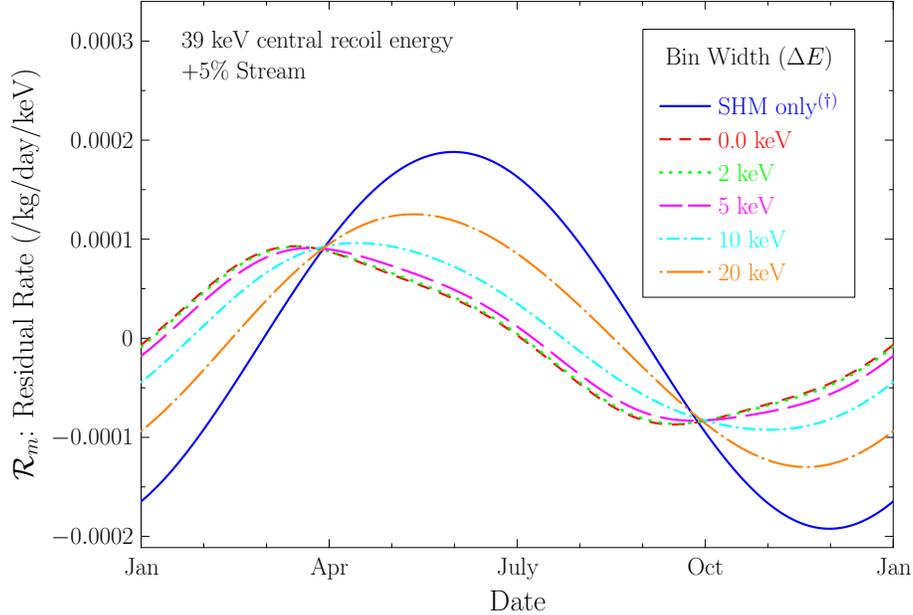}
  \caption[Recoil Energy Range]{
    \textit{Dependence on binning of recoil energy.}
    The average modulation observed over an energy range centered at
    39 keV ($E_{c,\Sgr}$) with a width $\Delta E$.  The solid (blue)
    line is for the SHM only with zero bin width $(\dag)$; the remaining
    lines include a 5\% Sgr stream in addition to the SHM.  Only for
    large bins, $\Delta E \gae 10$ keV, does the shape of the modulation
    differ significantly from that of the differential recoil rate
    (equivalent to zero bin width: $\Delta E = 0$  keV).
    }
  \label{fig:RecoilEnergyRange}
\end{figure}

We examine how the modulation appears when averaged over some range of
recoil energies, as would be observed when experimental data is binned,
and demonstrate the results in \reffig{RecoilEnergyRange}.  We have seen
that the Sgr stream has a significant effect on the modulation at least
over the recoil energy range 34-44 keV (see \reffig{RecoilEnergy}).
As detectors have limited energy resolution, experimental results are
often binned in recoil energies.  If the bins are much larger than the
limited width of the stream's $\orderof{1}$ relative modulation effect,
it would be difficult to study this behavior or even observe it.  In
\reffig{RecoilEnergyRange}, the modulation from the SHM and a 5\% Sgr
stream is shown for multiple bin sizes ($\Delta E$), where the bins are
centered at 39 keV, the characteristic energy $E_{c,\Sgr}$
(corresponding the energy cutoff; see \reffig{Eta}).  The average recoil
rate,
\begin{equation} \label{eqn:BinRate}
  \langle\dRdE\rangle = \frac{1}{\Delta E}
                        \int_{E_c - \Delta E/2}^{E_c + \Delta E/2}
                         \dRdE(E) \, \rmd E \, ,
\end{equation}
is used in each case.  For comparison, the SHM component alone is shown
at 39 keV.

An infinite resolution detector, with $\Delta E$ = 0, will observe the
actual recoil rate at $E_{c,\Sgr}$.  For bin widths of 2 \& 5 keV,
there is very little deterioration in the modulation signal.
Even for a 10 keV bin, the modulation signal is similar to the
$\Delta E$ = 0 case.  For $\Delta E$ = 20 keV, the signal is highly
deteriorated, but the modulation observed is still very different from
the SHM case.  The current generation of Ge detectors have energy
resolutions on the order of 1-2 keV, so it is clear detector resolution
will not inhibit observation of the stream's effect on the modulation
around $E_{c,\Sgr}$.

\subsection{\label{sec:Density} Stream Density}

\begin{figure}
  \insertfig{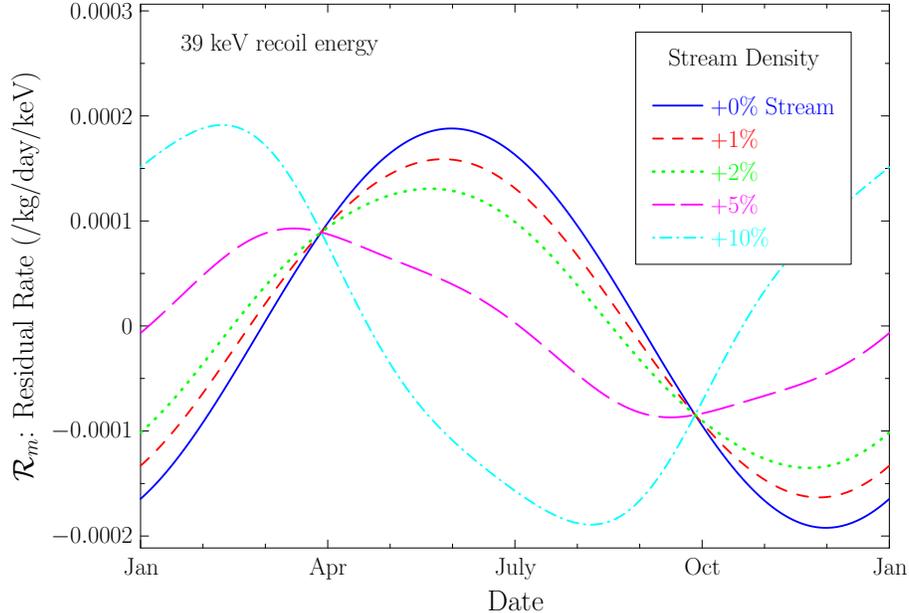}
  \caption[Stream Density]{
    \textit{Dependence on stream density.}
    The modulation for a variety of stream densities at a recoil energy
    of 39 keV ($E_{c,\Sgr}$); stream densities are given relative to the
    (fixed) SHM density.  The solid (blue) line represents the SHM
    alone.  Even at only 5\% of the density, the stream has a comparable
    modulation amplitude to that of the SHM alone near the cutoff
    energy, yielding a total modulation quite different than would be
    expected for the SHM alone.
    }
  \label{fig:Density}
\end{figure}

The extent to which a stream affects the annual modulation depends upon
the density of the stream.  The modulation due to the SHM + Sgr stream
is shown in \reffig{Density} for several different stream densities,
all at the characteristic energy $E_{c,\Sgr}$.  For comparison, the SHM
only case is shown by the solid (blue) line.  At this energy, the
modulation of the SHM peaks on June 1; that of the Sgr stream is
minimized on June 29.  The stream obviously significantly contributes to
the overall modulation at this energy for the 5\% relative density we
have been using.

While a 5\% density may be reasonable for some streams
\cite{Stiff:2001dq} (the caustic ring model of P.~Sikivie and
collaborators predicts a stream as large as 75\% of the local density;
see Ref.~\cite{Ling:2004aj} and references therein), it is
optimistically large for the Sgr stream.
Instead, a density of a few percent or lower may be more reasonable.
Would such low densities be observable?  In the figure, streams at 5\%
and 10\% densities differ significantly from the SHM component.  In
these cases, the modulation is non-cosinusoidal and asymmetric and has a
different amplitude and peak date than the SHM component.  For the 5\%
case, the maximum rate occurs 3 months prior to that of the SHM
component.  For a 2\% stream, the modulation shape and peak date do not
differ significantly from that of the SHM.  However, even though the
stream density is only $1/50$ that of the SHM, the amplitude of the
total modulation is still decreased by more than 30\% from that of the
SHM alone near $E_{c,\Sgr}$.  The implication then is, due to the
$\orderof{1}$ relative modulation of the stream near $E_{c,\Sgr}$, a
relatively small stream can still have a significant effect on the
experimental results.

\subsection{\label{sec:Dispersion} Stream Dispersion}

\begin{figure}
  \insertfig{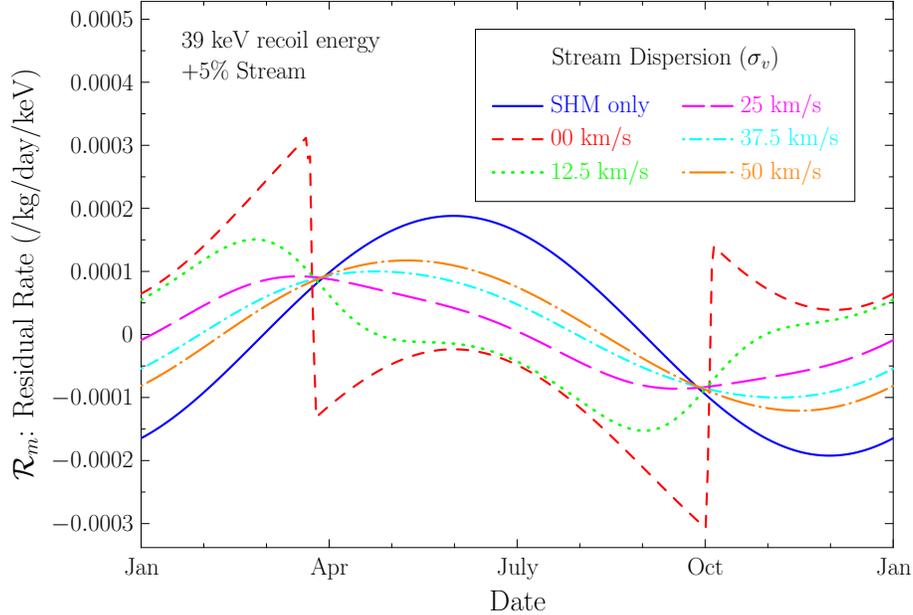}
  \caption[Stream Dispersion]{
    \textit{Dependence on stream dispersion.}
    The modulation at ($E_{c,\Sgr}$) is shown for various stream
    velocity dispersions ($\sigma_v$).  The solid (blue) line is for the
    SHM only; the remaining lines include a 5\% Sgr stream in addition
    to the SHM.  For $\sigma_v = 0$ km/s, the stream has a step function
    contribution on top of the cosine-like modulation of the SHM.
    Increasing velocity dispersion softens the step function like
    behavior.
    }
  \label{fig:Dispersion}
\end{figure}

We examine the dependence of the modulation on the velocity dispersion
$\sigma_v$ of the stream, illustrated in \reffig{Dispersion}.  Up to
this point we have taken the dispersion to be 25~km/s, and we now allow
it to vary.  The SHM and Sgr stream have very different modulation
behaviors: the SHM modulation is always very nearly cosinusoidal, with a
small amplitude and a reversal of the modulation phase at an energy
$E_{p,\SHM}$.  The stream modulation, on the other hand, has three
different behaviors: it is cosinusoidal with a small amplitude at
energies below the characteristic energy $E_{c,\Sgr}$, it is large
($\orderof{1}$ relative to the average rate) and non-cosinusoidal around
$E_{c,\Sgr}$, and it is non-existent above this energy.  The differences
between these two components are mainly due to the large difference in
their velocity dispersions, leading to the following two consequences:
(1) For a dispersion of a WIMP population significantly smaller than the
net motion $\vobs$ of that population relative to a detector, or
$\sigma_v \ll v_\odot$, the recoil energy spectrum develops a relatively
rapid drop near the characteristic energy $E_c$ (true for the Sgr stream
but not for the SHM).  (2) For a dispersion on the order of the
variation in $\vobs(t)$ due to the Earth's orbital motion, or 
$\sigma_v \lae b V_\oplus$, the variation of the location of this
dropoff leads to relatively large, non-cosinusoidal modulations in the
count rates near $E_c$ (again, this applies to the Sgr stream but not
the SHM).  This second condition essentially requires that the shift in
the location of the step ($\sim b V_\oplus$) be comparable or larger
than the width of the step dropoff itself ($\sim \sigma_v$); if this
condition is not satisfied, the modulation near $E_c$ remains a
relatively small effect (only a few percent of the total rate, as
occurs at other energies).
While the presence of a dropoff is signified by the condition
of (1), the (usually) stronger condition of (2) is necessary to make the
dropoff \textit{rapid enough} to observe the large modulation effect.

For the stream, $\eta(E)$ is similar to a step function, with the
position of the step, \refeqn{EcoSgr}, shifting in time; see
\reffig{Eta}.  The shift in the step is proportional to $b V_\oplus$,
the change in stream velocity $\vobs(t)$ (recall $0 \le b \le 1$ is a
geometrical factor dependent upon the stream direction).  The edge of
the step, however, is softened, with the fall off occurring over an
energy range proportional to $\sigma_v$.  As long as the velocity
dispersion is not significantly larger than the changes in $\vobs(t)$
(\ie\ $\sigma_v \lae b V_\oplus$), the edge is sharp enough so that, at
some energies, a large portion of the edge crosses the given energy
throughout the year.  Note how in \reffig{Eta}, $\eta$ is nearly atop
the step at 39 keV on December 28 ($t_{c,\Sgr}$) for the Sgr stream, but
is beyond the step on June 29. When $\sigma_v \gg b V_\oplus$, the step
is washed out, as can be seen by $\eta$ for the SHM in the same figure.
The small dispersion that is a characteristic of streams leads to the
$\orderof{1}$ relative modulation (arising from the crossing of the
step) that is not present in the SHM with its large dispersion.

The SHM + 5\% Sgr stream modulation is shown in \reffig{Dispersion} for
multiple stream dispersions at a recoil energy of 39 keV ($E_{c,\Sgr}$).
For comparison, the SHM only case is again shown by the solid (blue)
line.  The step like behavior in $\eta$ for the stream is apparent as
the dispersion goes to zero.  At $\sigma_v=0$, where $\eta$ is a true step
function, the step edge, $\Eco$, can be seen to have cross 39 keV in
late March and early October.  From March to October, the stream cutoff
energy is below the recoil energy being observed, so that the stream
does not contribute to the count rate (note how the shape of the
modulation on these dates is identical to the SHM only component).  From
October to April, the cutoff energy is above the recoil energy being
observed, so the stream also contributes to the count rate.

As $\sigma_v$ increases, the sharp changes in the modulation in March
and October are softened.  $\sigma_v$ = 12.5 km/s still looks somewhat
similar to the $\sigma_v=0$ case.  The $\sigma_v$ = 25, 37.5, \& 50 km/s
cases all have $\sigma_v$ greater than $b V_\oplus$ = 16 km/s; the
abrupt changes in the recoil rate are now gone.  However, $\sigma_v$ is
not significantly larger than $b V_\oplus$ and an appreciable portion of
the $\eta$ step still crosses this recoil energy: the lighter stream
still has an effect on the modulation comparable to that of the SHM, as
evident by the large difference between these cases and the SHM only
case.

As noted by Refs.~\cite{Stiff:2001dq,Helmi:1999ks,Helmi:1999uj},
the stream's velocity dispersion is more likely to be anisotropic,
in which case the distribution of \refeqn{Maxwellian} is not valid.
However, including anisotropic models of the stream's velocity
distribution would not significantly affect the results discussed in
this paper as the velocity dispersion of the stream will still be
significantly smaller than the stream velocity relative to the observer.
In such models, the dropoff in the recoil rate would still be present,
as well as a modulation of both the recoil rate and location of the
dropoff.  Including such models would simply lead to modest changes in
the shape of the dropoff in the mean inverse speed (see \reffig{Eta})
and, hence, the recoil rate around the characteristic energy $E_c$.

\section{\label{sec:Streams} General Streams}

Having examined various stream parameters using the Sgr stream in the
previous section, we wish to expand our results to streams in general.
The Sgr stream modulation features previously discussed likewise occur
for other streams:
(1) The phase of the stream modulation is associated with a
  characteristic time $t_c$ (typically the time of year at which the
  recoil rate is maximized or minimized) that is independent of the SHM;
(2) There exists a characteristic energy $E_c$ associated with a
  rapid dropoff to zero in the recoil rate;
(3) At energies less than the characteristic energy $E_c$, the
  modulation is nearly cosinusoidal, with a minimum at $t_c$ and a
  (small) amplitude of $\orderof{\frac{V_\oplus}{\vobs}}$ with respect
  to the average rate; and
(4) The modulation near $E_c$ is not cosinusoidal, is of $\orderof{1}$
  relative to the average rate, and can be comparable in amplitude to
  that of the overall halo modulation.
While any stream will have these features (we are assuming a small
velocity dispersion, or $\sigma_v \lae V_\oplus$, as would be expected
for typical streams), the characteristic time $t_c$ and energy $E_c$
depend upon the stream.  Given a stream velocity $\bv_{\odot,\Str}$,
these values may be determined from \refeqn{tc} and \refeqn{EcSgr},
respectively (using $v_{\odot,\Str}$ in place of $v_{\odot,\Sgr}$ in the
latter equation).  We will examine how the stream speed and direction
affect the characteristic time and characteristic energy, and
consequently the modulation.

\subsection{\label{sec:Speed} Stream Speed}

\begin{figure}
  \insertfig{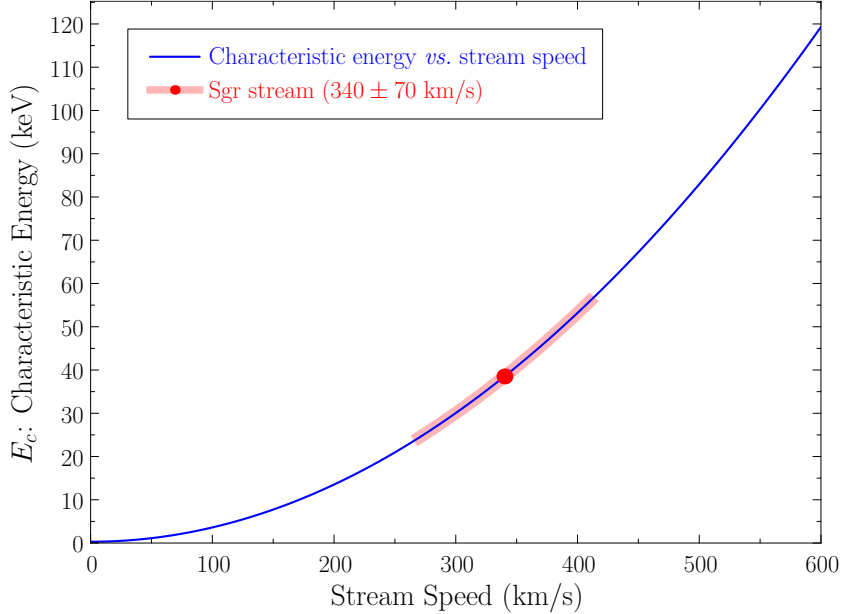}
  \caption[Stream Speed]{
    \textit{Stream speed.}
    The characteristic energy $E_c$, defined by \refeqn{EcSgr}, as a
    function of stream speed.  $E_c$ is approximately the highest
    possible recoil energy from a WIMP in the stream.
    $E_c$ is independent of the stream direction and depends only on
    the speed of the stream relative to the Sun ($v_{\odot,\Str}$;
    solid/blue line).  The $340 \pm 70$ km/s speed of the Sgr stream
    is shown (red dot with wide pink line indicating the $1\sigma$
    region), corresponding to an $E_c$ range of 25-55 keV with central
    value 39 keV.
    }
  \label{fig:Speed}
\end{figure}

The characteristic energy $E_c$ of a stream, \refeqn{EcSgr} with
$v_{\odot,\Sgr}$ replaced with a more general stream speed
$v_{\odot,\Str}$, is dependent only upon the speed of the stream
relative to the Sun and is independent of the stream direction.  For a
stream moving much faster than Earth's orbital velocity
($v_{\odot,\Str} \gg V_\oplus$), the characteristic energy is
proportional to the square of the stream speed relative to the Sun
($E_c \propto v_{\odot,\Str}^2$), so faster streams rapidly lead to
higher $E_c$'s; this behavior is demonstrated in \reffig{Speed}.

The $E_c$ determined from a modulation signal can be used to determine
the speed of the stream to an improved degree over that afforded by
limited alternative observations.  The $300 \pm 90$ km/s Sgr stream
speed estimate relative to the galactic rest frame, discussed in
Refs.~\cite{Majewski:2003ux,Freese:2003tt}, is not based upon direct
observation of Sgr stellar material in the local neighborhood, but
on extrapolating stellar stream observations in other areas of the
Milky Way, leading to the relatively large uncertainty.  The
$300 \pm 90$ km/s speed relative to the galactic rest frame corresponds
to a speed relative to the Sun of $v_{\odot,\Sgr} = 340 \pm 70$ and,
hence, to a 25-55 keV range of $E_c$, as shown in \reffig{Speed}.
This energy range is much larger than the energy resolution of a WIMP
detector; current detector resolutions of $\sim$1 keV would yield a
stream speed accurate to about $\pm$10 km/s (in the Sgr stream case).
To derive the stream speed from $E_c$, using \refeqn{EcSgr} (where $M$
is the nuclear target mass and $\mu$ is the WIMP-nucleus reduced mass),
the WIMP mass must already be reasonably well known.

\subsubsection{\label{sec:SpeedMass} Determining the WIMP Mass}

Alternatively, if
the WIMP mass is unknown, but a dropoff in the recoil rate can be
associated with a known stream (known via other observations), the
energy at which that dropoff occurs, $E_c$, can be used to determine the
WIMP mass.  For instance, if we observe a dropoff at $E_c = 30 \pm 2$
keV in a Germanium detector and associate it with the Sgr stream (with
speed $340 \pm 70$ km/s), we would derive a WIMP mass of $48 \pm 22$
GeV.  If further stellar observations and modelling of the Sgr stream
improve the speed uncertainties to $\pm 20$ km/s, the mass determination
would improve to $48 \pm 7$ GeV.  Certainly, the presence of a dropoff
in the recoil rate, which does not require a modulation effect to be
observed, can provide a useful tool for determining the WIMP mass.
However, care must be taken in associating such a dropoff with a
specific stream: observation of a 5\% density stream through such a
dropoff alone cannot be assumed to be due to the Sgr stream rather than
some as yet unknown other stream.  If we can determine the direction of
the stream producing the dropoff in the recoil rate, we have a much
stronger basis for associating that stream with a known one.  We examine
the stream direction in the following section.

\subsection{\label{sec:Direction} Stream Direction}

\begin{figure}
  \insertfig{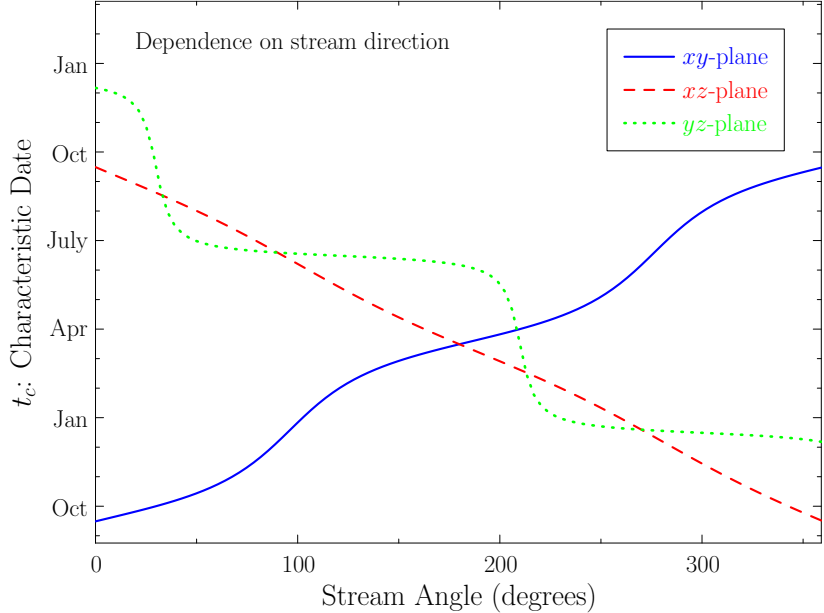}
  \caption[Stream Direction]{
    \textit{Stream direction.}
    The characteristic time $t_c$ as a function of stream direction;
    $t_c$ is independent of the stream speed.  Each curve shows the
    dependence on one of the three angles, in the $xy$-, $xz$-, and
    $yz$-planes (\eg, the angle $\phi$ in the $xy$-plane corresponds
    to the direction $\cos\phi \mathbf{\hat{x}}
    + \sin\phi \mathbf{\hat{y}}$ in Galactic coordinates).
    }
  \label{fig:Direction}
\end{figure}

The direction of a stream as well as the speed may be unknown to us.
Indeed, a signal in a dark matter detector may even be our first
indication of some local stream.  The characteristic time $t_c$ is
dependent only upon the stream direction, via \refeqn{tc}, so an
experimental determination of the former can give us indications of the
latter.  However, $t_c$ alone can only give the direction of the stream
in the Earth's orbital plane about the Sun, insufficient to reconstruct
the full stream direction.

The angle of the stream $\lambda_\odot$ from the normal of the orbital
plane is encoded in the geometrical factor $b$ discussed in
\refsec{EarthMotion}, with $b = \sin \lambda_\odot$.  The parameter $b$
can be determined only through modulation
effects due to the dependence of $\vobs(t)$, \refeqn{vobst}, on the
value of $b$.  A small value for $b$, corresponding to a stream nearly
orthogonal to the orbital plane, yields only small variations in $\vobs$
over the course of a year and modulation effects would likewise be
small; for a stream perfectly orthogonal to the orbital plane, $\vobs$
is constant ($b=0$) and there is no modulation.  A large value for $b$,
corresponding to a stream incident to the Sun along the Earth's orbital
plane, yields larger variation of $\vobs$ and larger modulation effects.
The dependence of the modulation on $b$ manifests itself primarily
through two effects: the modulation amplitude and the size of
the variation in the cutoff energy $\Ecomax-\Ecomin$ (see
\refeqn{EcoSgr}), both proportional to $b$.  The modulation amplitude is
degenerately dependent upon the presumably unknown stream density.  The
quantity $\Ecomax-\Ecomin$ is not, so $b$ can be extracted from a
modulation signal (the variation in $\Eco$ is also dependent upon
$v_\odot$, which may be unambiguously determined from the cutoff energy
as discussed in the previous section).

From $t_c$ and $b$, the full direction of the stream may then be
determined.  In \reffig{Direction}, we show $t_c$ as a function of angle
in the $xy$-, $xz$-, and $yz$-planes (angle is from the first axis
indicated in each plane).  The characteristic time $t_c$ varies most
smoothly with angle when the plane is near to that of the orbital plane
(the $xz$-plane is the nearest of the three shown).  The characteristic
time $t_c$ changes rapidly in the $yz$-plane (near 30$^\circ$
and 210$^\circ$) when the stream passes near the normal to the orbital
plane.  The otherwise flat behavior of $t_c$ in this case demonstrates
how $t_c$ is independent of the angle $\lambda_\odot$, as the near
orthogonality of the $yz$-plane with the orbital plane means rotations
in the $yz$-plane correspond strictly to changes in $\lambda_\odot$ and
not the angle in the orbital plane (except for a 180$^\circ$ phase shift
when rotating through the normal to the orbital plane).

The determination of the parameters $t_c$ and $b$ from a detector signal
is essentially independent of the WIMP mass.  Then, even without an
understanding of WIMP properties (\eg\ mass), we can use these two
parameters (by converting them to a stream direction) to associate a
dark matter detector signal with possibly visible structure in the
galaxy, such as a stellar stream, that is moving in the same directon.
Without knowing the WIMP mass or the direction of the stream producing
the signal in the detector, a dropoff in the recoil rate spectrum at a
characteristic energy $E_c$ cannot be associated with any specific
known structure (known via other observations), \eg\ the Sgr stream,
rather than some other unknown stream.  As described in the previous
section, any visible structure could give an indication of the speed of
the dark matter stream and, therefore, lead to a determination of the
WIMP mass by the position of the dropoff via \refeqn{EcSgr}.

\subsection{\label{sec:Parameters} Determination of Stream Parameters}

As noted by Stiff, Widrow \& Frieman \cite{Stiff:2001dq}, the modulation
signal can be used to determine many of the characteristics of such a
stream; we note here that such determinations can be made mainly from
a small energy range about $E_c$ and, in the following, outline how to
do so.  The approximate energy of the rapid change in modulation
behavior, corresponding to $E_c$, yields the stream speed
$v_{\odot,\Str}$ since $E_c \propto v_{\odot,\Str}^2$.  The location of
this cutoff energy varies over the course of a year; the extent of the
variation depends upon the geometrical parameter $b$ (the sine of the
angle between the stream and the normal to Earth's orbital plane), which
can thus be extracted.  From $b$ and the characteristic date $t_c$
(the time of year at which the stream moves fastest relative to the
Earth; determined from the date of the peak rate), the direction of
the stream can be determined via the formulas of \refsec{EarthMotion}.
The amplitude of the modulation can be used to determine the density of
the stream.  The width of the cutoff indicates the velocity dispersion
in the stream.  While observation of these characteristics could
potentially be inhibited by limited energy resolutions in dark matter
detectors, the resolutions available in some current detectors should be
sufficiently high as to not pose a significant impediment to extracting
many of these stream parameters.

The various parameters, while all potentially extractable from an
observed modulation signal, will not be equally easy to extract.
The characteristic energy $E_c$, characteristic time $t_c$, and stream
density $\rho_\Str$ relative to the smooth halo density (assuming the
halo modulation is observed at energies not near $E_c$) are the
parameters most easily extracted from any modulation signal as they
have the most dominant effect on such a signal.  From the first two of
these parameters, two components of the stream velocity can be
determined.  The stream dispersion $\sigma_\Str$ and the geometrical
factor $b$ would be more difficult to extract as they have less
significant effects on the modulation signal; extracting these
parameters might require much more significant detector exposure and/or
a larger detector.  The stream dispersion will give an indication of the
velocity distribution of WIMPS in the stream, but the interpretation of
this parameter should be limited to an approximation of the distribution
only as our assumed Maxwellian distribution of \refeqn{Maxwellian} may
not be entirely accurate (see the discussion at the end of
\refsec{Dispersion}).  Indeed, directional detectors are
likely to be much more useful in characterizing this velocity
distribution.  From $b$, the third component of the stream velocity
can be determined.  Alternatively, if a stream is due to a dwarf galaxy
being absorbed by the Milky Way, there could be an associated stellar
stream that would be indicative of the WIMP stream, yielding independent
measurements of the stream velocity.  However, for late infall of dark
matter clumps and other possible WIMP stream origins, there would be no
such stellar stream and WIMP detection alone would be required to
characterize the stream parameters.

A stream signal in a dark matter detector may provide a means of
determining the WIMP mass, as discussed previously.  Individual
parameters that play a role in this determination are discussed here.
The parameters $E_c$, $t_c$, and $b$ can be determined from a detector
signal without knowing the mass of the WIMP.  The latter two parameters
would give an indication of the direction of the dark matter stream
producing the signal.  Knowing this direction, we could associate the
dark matter stream with some stellar stream or known structure in the
galaxy, such as the Sgr stream, and independently determine the speed
of the dark matter stream (which presumably matches that of any visible,
\eg\ stellar, components).  The characteristic energy $E_c$ depends only
upon the WIMP mass, nuclear target mass, and stream speed, so by
knowing the velocity of the stream producing a dropoff in the recoil
rate at $E_c$, the WIMP mass can be derived (the nuclear target mass is
well known in most detectors).  The mass determination from a stream in
this manner could be more precise than that afforded by the signal from
the smooth halo itself.

Detectors that can determine the direction of WIMP induced nuclear
recoils are the ultimate goal for characterizing the local WIMP
population; however, technology for such a detector is still in the
development stage.  In the meantime, probing the annual modulation will
also allow us to characterize this WIMP population by, \eg, observing
the presence of streams in addition to the smooth dark matter halo.

\section{\label{sec:Summary} Summary}

A dark matter stream presents observational signals unlike that of a
smooth background distribution (assumed here to be the SHM).  The most
noticeable difference is the existence of a rapid dropoff in the recoil
rate around a characteristic energy $E_c$; this dropoff yields a
relatively large and non- cosine-like contribution to the annual
modulation around $E_c$ that may be observable even for a stream much
less dense than the SHM.  \refFig{Components} shows the drastic
importance the stream can have near $E_c$ for the annual modulation.
The presence of a rapid change in modulation behavior with respect to
recoil energy would be an indication of a stream or other non-SHM
component of the dark matter.  Since a stream's effects are mild except
near $E_c$ (and essentially non-existent well above $E_c$), the presence
of a stream should not interfere with using the modulation to describe
the background distribution, SHM or otherwise.  Detection of an annual
modulation of the cutoff energy together with the annual modulation
of the overall signal provide a ``smoking'' gun for WIMP detection.
In addition,  if the WIMP mass is unknown, but a dropoff in the recoil
rate can be associated with a known stream (known via other
observations), the energy at which that dropoff occurs can be used to
determine the WIMP mass.


\begin{acknowledgments}
  C.S.\ and K.F.\ acknowledge the support of the DOE and the Michigan
  Center for Theoretical Physics via the University of Michigan.
  P.G.'s work was partially supported by NSF grant PHY-0456825.
  C.S.\ thanks B.\ Burrington for providing comments on this paper.
  C.S.\ thanks the unknown audience member at the UCLA Symposium on
  dark matter (Marina del Rey, Feb 2006) who raised the question about
  the possibility of streams providing a mass determination for the
  WIMP, leading to the inclusion in this paper of how to do so.
  We thank H.\ Newberg for useful comments.
\end{acknowledgments}




\end{document}